\documentclass{aa}
\usepackage{graphicx}
\usepackage{epsfig}
\usepackage{graphics}
\begin{document}%
   \title{Very high energy $\gamma$-ray emission from X-ray transients during major outbursts}

   \author{M. Orellana\inst{1,2}\thanks{Fellow of CONICET, Argentina},
           G.~E. Romero\inst{1,2}\thanks{Member of CONICET, Argentina},
           L.~J. Pellizza\inst{3}
           \and S. Vidrih\inst{4}
          }

   \offprints{M. Orellana {\em morellana@fcaglp.unlp.edu.ar}}
   \titlerunning{VHE $\gamma$-ray emission from X-ray transients}

\authorrunning{M. Orellana et al.}

\institute{Facultad de Ciencias Astron\'omicas y Geof\'{\i}sicas, Universidad Nacional de La Plata, Paseo del Bosque, 1900 La Plata, Argentina \and Instituto Argentino de Radioastronom\'{\i}a, C.C.5,
(1894) Villa Elisa, Buenos Aires, Argentina \and Instituto de Astronom\'{\i}a y F\'{\i}sica del Espacio, C.C. 67, Suc. 28, 1428, Buenos Aires, Argentina\and Institute of Astronomy, University of Cambridge, Madingley Road, Cambridge CB3 0HA, United Kingdom}

\date{Received / Accepted}


  \abstract
   {Some high mass X-ray binaries (HMXB) have been recently confirmed as $\gamma$-ray sources by ground based Cherenkov telescopes. In this work, we discuss the $\gamma$-ray emission from X-ray transient sources formed by a Be star and a highly magnetized neutron star. This kind of systems can produce variable hadronic $\gamma$-ray emission through the mechanism proposed by Cheng and Ruderman, where a proton beam accelerated in the pulsar magnetosphere impacts the transient accretion disk. We choose as case of study the best known system of this class: A0535+26.}
   {We aim at making quantitative predictions about the very high-energy radiation generated in Be-X ray binary systems with strongly magnetized neutron stars.}
   {We study the gamma-ray emission generated during a major X-ray outburst of a HMXB adopting for the model the  parameters of
A0535+26. The emerging photon signal from the disk is determined by the grammage of the disk that modulates the optical depth. The
electromagnetic cascades initiated by photons absorbed in the disk are explored, making use of the so-called ``Approximation A''
to solve the cascade equations. Very high energy photons induce Inverse Compton cascades in the photon field of the massive star. We implemented Monte Carlo simulations of these cascades, in order to estimate the characteristics of the resulting spectrum.}
  {TeV emission should be detectable by Cherenkov telescopes during a major X-ray outburst of a binary formed by a Be star and a highly magnetized neutron star. The $\gamma$-ray light curve is found to evolve in anti-correlation with the X-ray
signal.}
   {}


   \maketitle
%
\section{Introduction}
Galactic transient and variable X-ray sources have long been proposed as $\gamma$-ray emitters (see e.g. Rudak \& M\'esz\'aros 1991, Cheng et al. 1991, Anchordoqui et al. 2003). 
In fact, some sources observed by the Energetic Gamma Ray Experiment Telescope (EGRET) have already been suggested to be associated with high mass X-ray binaries (HMXBs): A0535+26 (Romero et al. 2001),
LS~I~+61$\degr$303 (Kniffen et al. 1997), Cyg X-3 (Mori et al. 1997),  LS 5039 (Paredes et al. 2000). Moreover, very high energy $\gamma$-ray emission was recently detected from the microquasars LS 5039 (Aharonian et al. 2005a) and LS~I~+61$\degr$303 (Albert et al. 2006), and from the colliding wind Be-pulsar system PSR B1259-63 (Aharonian et al. 2005b).

In accreting HMXBs the matter is usually transferred onto the compact object via the stellar wind of the massive star. In addition to its regular mass loss, the  primary, commonly of Be spectral type, has recurrent phases of violent ejection. This ejected material arrives to the vicinity of the compact object giving rise to enhanced X-ray emission (an X-ray outburst).
In some cases, the HMXB contains a strongly magnetized neutron star (NS), being the magnetic field intense enough as to disrupt the accretion disk. The accreting material is then channeled through the field lines onto the surface of the NS (e.g. Gosh \& Lamb, 1979).
The Cheng-Ruderman mechanism acts in the magnetosphere of such an accreting NS to accelerate protons up to relativistic (multi-TeV) energies (see a description of  the general picture in Orellana \& Romero 2005, and
references therein). These energetic protons move along the magnetic field lines to impact into the accretion disk. Their interaction with the material of the disk initiates hadronic and electromagnetic showers.

In order to make quantitative predictions about the high-energy emission originated in this particular kind of systems we have adopted as case of study A0535+26, which is the best studied among the transient Be/X-ray binaries. A major outburst of this system was observed in 1994 with the instrument BATSE of the Compton Gamma-Ray Observatory (Finger et al. 1996).
Its light curve allows to estimate the history of the neutron star mass accretion rate, which is used in this work to calculate the evolution of the parameters that determine the high energy emission. The power injected inside the disk by $\gamma$-rays resulting from neutral pion decays can be then estimated. We show that the cascades triggered in the disk are dominated by Bremsstrahlung cooling of the electrons. In a broad energy range the development of these cascades can be estimated by solving the cascade equations applying an analytical approach, the so-called {\em Approximation A} (Rossi \& Greissen, 1941). The gamma-ray signal emerging from the other side of the disk is formed by the unabsorbed photons and by the cascade recycled radiation.

The propagation of the gamma-rays in the next stage is determined by their interaction with the radiation field of the donor star. Detailed analysis of the anisotropic cascades inside massive binary systems have been performed by
Bednarek (1997, 2000, 2006) and Sierpowska \& Bednarek (2005). We have simulated the Inverse Compton (IC) cascades under some simplifying assumptions. For this purpose we have implemented a code based on the scheme presented by Protheroe (1986) and Protheroe et al. (1992), that allows us to estimate the emerging spectrum at high energies. Detectability of the TeV flux by the new generation of Cherenkov telescopes can be then discussed.

The structure of the paper is as follows: first, we examine how $\gamma$-rays can be produced in transient Be/X-ray binaries; Section 2 presents a brief review of the Cheng-Ruderman magnetospheric model and the gamma-ray generation in such a context. Section 3 provides some information on the archetypal system A0535+26. In Section 4 we present the results of our time-dependent calculations adopting the input parameters determined during a major outburst in this source. Sections 5 and 6 deal with the cascades induced in the disk by the relativistic protons, and by the emerging $\gamma$-rays in the ambient photosphere. We close in Section 7 with a discussion.


\section{Gamma-ray emission from HMXB containing a highly magnetized neutron star}
\subsection{The Cheng-Ruderman magnetosphere model}

\begin{figure}
\centering \psfig{figure=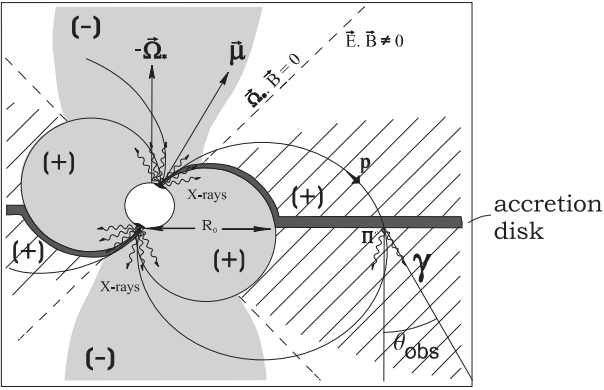, width=8.0cm}
\caption{Cheng-Ruderman mechanism at work in the magnetosphere of a HMXB containing a strongly magnetized neutron star (adapted from Romero {\it et al.} 2001)}\label{cheng}
\end{figure}

Cheng \& Ruderman (1989, 1991) have studied the magnetosphere of a
neutron star surrounded by a Keplerian accretion disk. When the
inner parts of the disk rotate faster than the star, inertial
effects lead to a magnetospheric charge separation. A gap entirely
empty of plasma is formed and it separates oppositely charged regions. In this gap
${\bf E\cdot B}\neq 0$ and a strong potential drop $\Delta V$ is
established. For strong shielding conditions in the gap the potential drop can reach the static gap value $\Delta V \sim 4 \times 10^{14}\beta^{-5/2}\left(\frac{M_{*}}{M_{\odot}}\right)^{1/7}\!\!
R_6^{-4/7}L_{37}^{5/7}B_{12}^{-3/7}$ V, where $M_{*} \sim 1.4$ $M_{\odot}$ is the mass of the neutron star, $R_6$
is its radius in units of $10^6$ cm, $B_{12}$ is the magnetic
field in units of $10^{12}$ G, and $L_{37}$ is the X-ray
luminosity in units of $10^{37}$ erg s$^{-1}$. The parameter
$\beta\equiv 2R_0/R_{\rm A} \sim 1$ is twice the ratio of the
inner accretion disk radius to the Alfv\'en radius. 
At a radius $r\simeq R_0$ the disk is disrupted by the magnetic pressure and the matter is channeled by the magnetic field lines to the neutron star polar cap, where it impacts producing hard X-ray emission. Cheng \& Ruderman (1989, 1991) have analyzed some possible configurations of the static gap for the aligned rotator, but a realistic three-dimensional model for the misaligned case $(\bar{\Omega}\times\bar{\mu}\ne 0)$ has not been developed yet.
For illustrative purposes, a sketch of the mechanism is presented in Figure \ref{cheng}. 

For typical parameters in Be/X-ray binaries, $\Delta V\sim
10^{13-14}$ V (Romero {\it et al.}, 2001). Protons entering into the gap
from the stellar co-rotating region will be accelerated to
multi-TeV energies and directed toward the accretion disk along the
magnetic field lines. The maximum current that can flow through the gap can be
determined from the requirement that the azimuthal magnetic field
induced by the current does not exceed that of the initial field
${\bf B}$:  $J_{\rm max}\sim B_{\star}R_{\star}^3 R_0^{-2}$, where $B_{\star}$ is the magnetic field and $R_\star$ is the radius of the neutron star (Cheng \& Ruderman, 1989).

Note that the impact of energetic protons on the disk does not occur only at a singular point but over a certain area that is radially extended, following the spread of the field lines that pervade the gap and are dragged by the disk. Some authors incorporate a beaming factor ($\Delta\Omega/4 \pi\approx 0.3$) in the escaping photon luminosity to take into account in phenomenological way of the anisotropy of the emerging $\gamma$-ray emission (Cheng et al. 1991, Romero et al. 1991).


The accretion disk is partially threaded by the magnetic field
lines of the NS. The screening effect of the currents
induced in the disk surface can be taken into account through a screening factor $\eta\le 1$. Then, the magnetic field is $B_z=\eta \mu /r^3$ inside the disk, where $\mu$
is the magnetic dipolar moment of the NS. An expression for the
location of the inner radius $R_0$ where the disk is disrupted by the magnetic forces, as a function of the screening factor, was found by Wang (1996).

An azimuthal component of the magnetic field is originated by the
relative rotation, and it is expected that $B_z\sim B_\phi$
close to $R_0$. In Orellana \& Romero (2005) two situations were considered concerning the
screening of the disk i.e. $\eta=$ 1 and 0.2. The way the
results of calculations are affected by this change was discussed in that paper.
Throughout the present work we consider $\eta=0.1$, for a partially
screened disk.

\subsection{$\gamma$-ray production}

Relativistic protons entering into the accretion disk interact with
target protons to produce secondary charged and neutral $\pi$-mesons.
The subsequent neutral pion decays
lead to gamma-ray production. Charged pions of multi-TeV energies
are able to interact with protons of the disk before the decay time (Anchordoqui {\it et al.} 2003).
Such interaction occurs with a cross section $\sigma_{\pi^\pm
p}\sim 22$ mb, in this energy range.

The differential $\gamma$-ray emissivity
is
\begin{equation}
q_{\gamma}(E_{\gamma})=2\int^{\infty}_{E_{\pi}^{\rm
min}(E_\gamma)} \frac{q_{\pi^0}(E_{\pi})}{\sqrt{E_{\pi}^{2}-
m_{\pi}^{2} c^4}} \;dE_{\pi}, \label{qgama}\end{equation} 
where $E_{\pi}^{\rm
min}(E_{\gamma})=E_{\gamma}+\frac{m_{\pi}^{2} c^4}{4E_{\gamma}}$. Applying the $\delta$-function approximation for the
 differential cross section\footnote{This approximation considers only the most energetic neutral pion that is produced in the $pp$ reaction aside of a {\em fireball} composed by a certain number 
 of less energetic $\pi$-mesons of each flavor. See a discussion in Pfrommer and En$\beta$lin (2004).}(Aharonian \& Atoyan, 2000), the pion emissivity becomes
\begin{eqnarray}
q_{\pi^0}(E_{\pi})&= 4 \pi \int_{E_{\rm th}}^{\infty}
\delta(E_\pi-\kappa{E_{\rm kin}}) J_p(E_p)\,\sigma_{pp}(E_p)
 \,dE_p\nonumber \\
 &=\frac{4\pi}{\kappa} J_p\left( m_p c^2+\frac{E_{\pi}}{\kappa}\right)\sigma_{pp}\left( m_p c^2+\frac{E_{\pi}}{\kappa}\right)
\end{eqnarray}
for proton energies greater than the energy threshold $E_{\rm
th}=1.22$ GeV. Here, $\kappa$ is the mean fraction of the
kinetic energy $E_{\rm kin}=E_p-m_p c^2$ of the proton transferred
to a secondary meson per collision. For a broad energy
region (GeV to TeV) $\kappa\sim 0.17$. The total cross
section of the inelastic pp collisions is well approximated by
$\sigma\sim 30\left(0.95+0.06 \ln (E_{\rm kin}/{\rm GeV})\right)$ mb.

\begin{figure}
\centering \psfig{figure=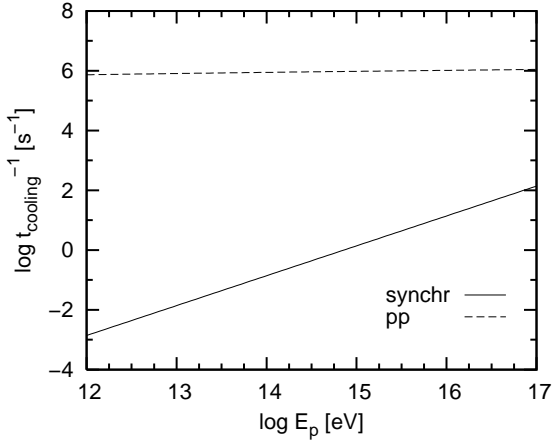, width=6.0cm,angle=-90}
\caption{Cooling rates of high-energy protons impacting the accretion disk. In the calculations we have adopted $B=2.5\times 10^6$ G and $n_p=1.2 \times 10^{21}$ cm$^{-3}$ for the magnetic field and proton density inside the disk, respectively.}\label{cool}
\end{figure}

In order to compare the relative importance of the proton energy losses in the accretion disk we have evaluated the cooling rates for $pp$ interactions and synchrotron radiation. The latter process can take place because of the magnetic field lines are strongly twisted in the accretion disk. Hence, for a proton, even if initially moving along the field line, its inertia can force it to depart from the line inside the disk, where it starts to emit synchrotron radiation. The cooling rates are given by (e.g. Begelman, Rudak \& Sikora, 1990)
\begin{equation}
\left[t_p^{\rm synchr}\right]^{-1}=\frac{\sigma_{\rm T} m_e B^2 E_p}{6 \pi m_p c^3}\label{tSy}
\end{equation}

\begin{equation}
\left[t_p^{pp}\right]^{-1}=K \sigma_{pp} c n_p,\label{tpp}
\end{equation}
where $K\simeq 0.5$ is the inelasticity (fraction of the proton energy lost per interaction), and $\sigma_{\rm T}$ is the Thomson cross section. For the particular case studied in the next sections we have that $t_p^{pp}\ll t_p^{\rm synchr}$ over a wide range of energy (see Figure \ref{cool}). In addition, it should be noted that the crossing time $t_{\rm cross}\sim2h/c$ of a proton traversing the disk is much longer than $t_p^{pp}$, so the protons effectively cool by $pp$ interactions in the disk, being the synchrotron losses negligible. 

Since the protons arrive to the disk with basically the same energy
acquired at the gap, estimated as $E_p=e\, \Delta V$, the
proton injected spectrum can be then estimated as
\begin{equation}
J_p (E) \sim \frac{J_{\rm max}}{e \pi R_0^2}\delta(E-E_p)\;\;\; \frac{\rm protons}{\rm s \;cm^2}. \label{J_p}
\end{equation}

Performing the integral in expression (\ref{qgama}) we obtain
\begin{eqnarray}
q_{\gamma}(E_{\gamma})&\approx & \frac{2\,J_{\rm max}}{e \pi R_0^2}\times\nonumber\\
&& \frac{\sigma_{pp}(E_p)}{\sqrt{\kappa^2(E_p-m_p c^2)^2-m_\pi^2 c^4}}\,{\rm
\frac{ph}{atom\,cm^3\,eV\,s}},\label{qug}\end{eqnarray}
for $E_{\gamma}$ between $E_1=0.5 E_\pi(1-v_\pi/c)$ and $E_2=0.5
E_\pi(1+v_\pi/c)$ and $q_{\gamma}(E_{\gamma})=0$ outside this range. Here $v_\pi$ is the velocity of the pion with
energy $E_\pi=\kappa (E_p-m_pc^2)$. The luminosity released in the
disk through $\pi^0$-decays results:

\begin{equation}
\label{Int} L^0(E_{\gamma})=\pi R_0^2 \int_0^{2h}
q_{\gamma}(E_{\gamma}){E_{\gamma}}^2 n_p\,
e^{-z/\lambda_{pp}}\;dz, \end{equation} where $n_p$ is the number density of
the protons in the disk, and $\lambda_{pp}$ is the mean free path
of a proton entering into the disk. Taking into account the opacity of
the disk (absorption in the Coulomb field of nuclei, for which $\sigma_{\gamma
p}\sim 10^{-26}$ cm) we can estimate the luminosity that escapes. Energetic photons that are absorbed initiate electromagnetic cascades inside the disk, which are discussed in
Section \ref{aprox-A}. The unabsorbed high-energy photons conserve their original spectrum,
completely dominated by the most energetic photons, as can be noted from
 (\ref{qug}) and (\ref{Int}). The maximum energy is given by
$E_\gamma^{\rm max}\simeq \kappa E_p$.

The photons escape along the original arrival direction of the primary protons. As it was mentioned in Section 2.1, the protons follow the field lines along the gap, which form different angles with the disk surface due to the differential dragging. This will produce some beaming of the emission, but not very strong, in such a way that the duty cycle will render a detectable source for favorable viewing conditions.

\section{A case of study: A0535+26}
Among the Be/X-ray binaries, A0535+26 has been noted for its
remarkable multi-frequency behavior. This motivated extensive
observational campaigns that have made of it the best known system
in its class, and a good subject for the model discussed here. A complete review of
A0535+26 is given by Giovannelli \& Graziati (1992).

A0535+26 is an eccentric system formed by a slow X-ray pulsar (104 s)
and the B0III-type star HDE 245770. The orbital period is 111 days. The system has been frequently observed to undergo
transient X-ray flares with a wide range of intensities. In
fact, very recently it underwent another major outburst that
was detected with instruments of the Rossi X-ray Timing Explorer
(Coe et al., 2006). The simultaneity of X-ray activity with
transitions between infrared states strongly suggests that these
outbursts occur in relation with increments of the accretion rate
onto the neutron star. The variations in the accretion rate are thought to be related to the presence of a transient circumstellar disk around the Be donor star (Haigh, Coe \& Fabregat, 2004).

The distance to A0535+26 is estimated in $2.6\pm 0.4$ kpc.
The spectral feature detected at X-ray energies near 110 keV can
be interpreted of cyclotron origin, indicating that the
neutron star has an intense polar magnetic field of $\sim
10^{13}$ Gauss (Finger {\it et al.} 1996).

Romero {\it et al.} (2001) suggested that the Cheng-Ruderman mechanism acting
in A0535+26 could be a plausible explanation for the gamma-ray source 3EG J0542+2610. In the next sections we will develop a more quantitative model for the possible $\gamma$-ray emission of A0535+26.

\section{Time dependent model}

A major outburst of A0535+26 was detected at hard X-rays (20-100
keV, reaching a peak flux of $\sim$8 crab) during February-March 1994, with no simultaneous observations at lower energies. A sequence of normal outbursts was detected
before and after this giant one (see Stella, White \& Rosner 1986, for a description of different outburst types). The peak of the hard
X-ray flux of the 1994 major outburst occurred after the
periastron passage. Analysis of the QPO features and of the
inferred spin-up history strongly supports the idea of the
formation of a transient accretion disk around the neutron star
(Finger et al. 1996). The hard X-ray emission detected is thought
to be originated through the collision of the accreted matter
onto the surface of the neutron star, at the polar caps.
The accretion disk would emit mostly at soft X-ray energies, likely with a higher flux.

We have fitted a smooth curve to the flux values reported by Finger et al.
(1996). At the peak of the outburst ($t=0$ in our notation) the disk luminosity was assumed to reach $L=10^{38}$ erg s$^{-1}$. This normalization results from an extrapolation of the flux detected by BATSE to lower energies ($\sim$ few keV). The
light curve was used to estimate the evolution of the relevant
physical parameters of the system, assuming $\epsilon\sim 10\%$
as the radiative efficiency. 
The radiation pressure is assumed to dominate over
other pressure components (magnetic field, gas). This fact determines the
structure of the inner region of a standard accretion disk, which is assumed to be partially threaded by the magnetic field of the NS. 

\begin{figure*}
\centering \psfig{figure=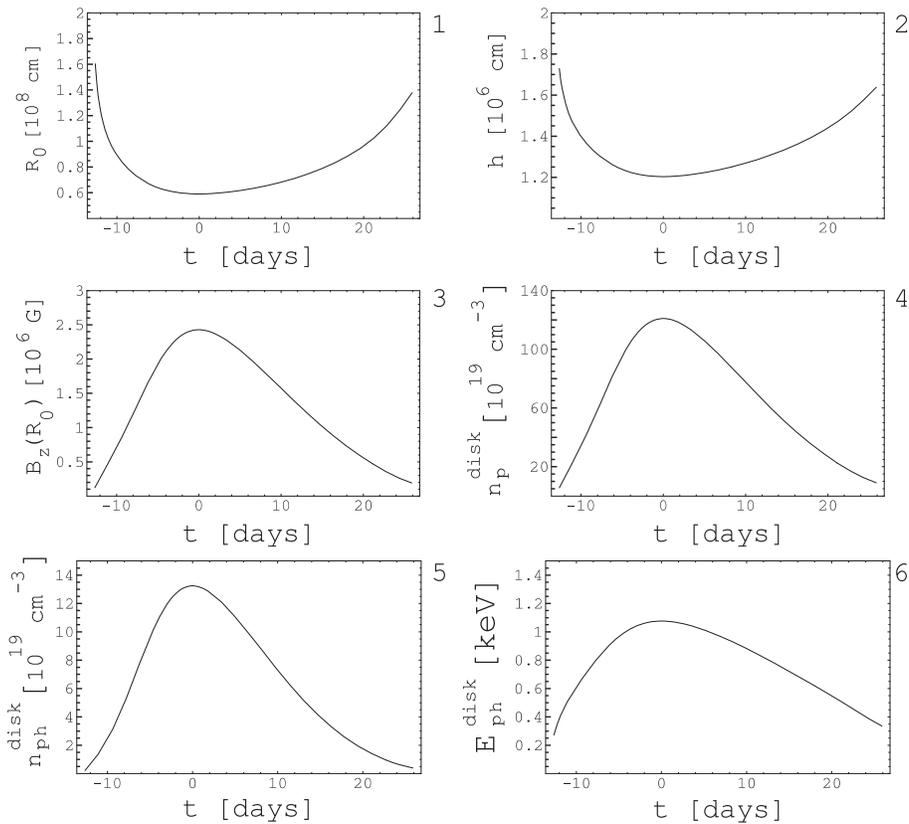, width=12.0cm}
\caption{Temporal evolution of the physical parameters of the accretion
disk, $t=0$ indicates the maximum of the X-ray outburst.  1)Inner radius $R_0$, where the disk is disrupted by the magnetic
forces. The following values are estimated at $R_0$. 2) Half
height of the accretion disk. 3) Vertical component of the
magnetic field inside the disk for the assumed screening factor.
4) Numerical density of protons inside the disk. 5) Numerical
density of the thermal photons emitted by the accreting matter. 6)
Characteristic energy of the mentioned photons. }\label{disk}
\end{figure*}

\begin{figure*}
\centering \psfig{figure=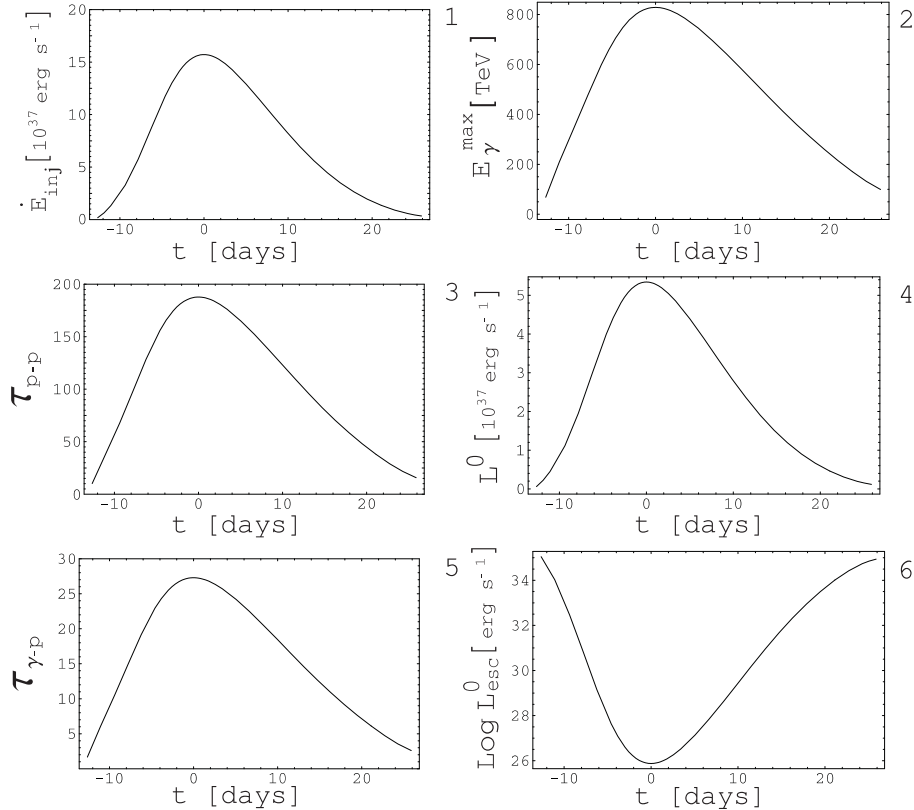, width=12.0cm}
\caption{Parameter evolution associated with the effects of proton impact onto the disk, and the luminosity generated after the pion decay.
1) Energy power injected in the disk by the proton bombardment,
2) Maximum energy of the gamma-ray photons resulting from the $\pi^0\rightarrow 2 \gamma$ decay,
3) Optical depth for a proton entering into the disk,
4) Luminosity generated through neutral pion decay,
5) Optical depth for the most energetic photons,
6) Emerging luminosity, estimated as $L^0 e^{-\tau_{\gamma p}}$. }\label{inicio-cas}
\end{figure*}

In the panels of Figure \ref{disk} and \ref{inicio-cas} we show the temporal  evolution of the inner radius of the accretion disk, the half-height of the disk, the $z$-component of the magnetic field, the particle density and the photon density inside the disk assuming black body emission, the average energy of such photons, the injected energy into the disk in the form of relativistic particles, the maximum energy of the first-generation gamma-rays inside the disk, the opacity of the disk to proton propagation, the gamma-ray luminosity produced inside the disk as a consequence of the $\pi^0$-decays, the opacity of the disk to these high-energy gamma-rays, and the luminosity of the gamma-rays that finally emerge from the other side of the disk. All values are given at $R_0$. The expressions used to perform these calculations can be found in Orellana \& Romero (2005). 
Figure \ref{cool} shows the comparison of the cooling rates given by (\ref{tSy}) and (\ref{tpp}), adopting the highest values for $B$ and $n_p$ reached during the outburst.

The total energy injected in the disk by the proton bombardment is
huge,
 as it is the power released through neutral-pion decays (i.e. $L^0\sim 10^{37}$
 erg s$^{-1}$).
The optical depth of the disk to the propagation of these photons
follows a time evolution similar to the rate of accretion of mass,
reaching $\tau_{\gamma\,p}\sim27$ at the peak of the X-ray outburst. Therefore at high energies a deep valley in the light curve is found, in anti-correlation
with the X-ray behavior. 


\section{Cascades inside the disk}\label{aprox-A}

The protons accelerated in the gap reach the disk following the magnetic
field lines. The particles involved in the further cascades move
nearly in the same direction of the initial protons. We consider a
longitudinal treatment of the cascade. It starts at the point of
the $\gamma-$photon production (after the $\pi^0$ decay) and develops
along the proton arrival direction. This direction is not well
determined since we have estimations for $B_z$ and can argue for
$B_\phi \sim B_z$ inside the disk, but the radial component of the
magnetic field $B_r$, which determines the $\theta_{\rm obs}$
angle, remains as a free parameter (see Figure \ref{cheng}). The
latter component of the magnetic field inside the disk depends on
the angle of misalignment between $\bar{\Omega}$ and $\bar{\mu}$ (the vector of the axis of rotation and the magnetic dipolar moment), and on the inward drag of the field lines by the conducting accreting matter. We follow the cascade inside the disk considering a pitch angle of $45$ degrees with the $z$
direction (due to $B_\phi \sim B_z$). 


The energy losses of the electrons that take part in the
cascades inside the disk, result from the interaction with the
surrounding ambient (see in Figure \ref{disk} the evolution of the
physical parameters of the disk). By comparing the cooling times
that characterize the rate of the relevant processes
(Bremsstrahlung and Inverse Compton) we determine which dominates.

\begin{figure}
\centering \psfig{figure=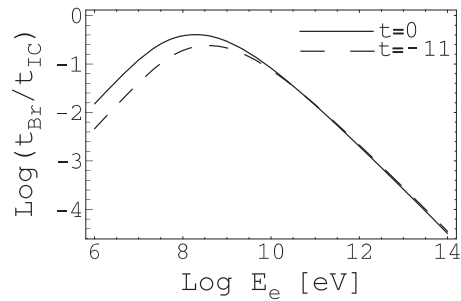, width=6.0cm}
\caption{ Ratio of cooling times for the main interactions of electrons in the accretion disk. The dominance of Bremsstrahlung allows the introduction of the simplifications of the so-called {\sl Approximation A} for the calculation
of the electromagnetic cascades inside the disk. Two
representative instances have been considered.}\label{dominaB}
\end{figure}

The ratio between these cooling times $(t_{\rm Br}/t_{\rm IC})$ is
shown in Figure \ref{dominaB} as a function of the electron energy.
Bremsstrahlung dominates during the entire outburst. Therefore we can apply
some simplifying hypothesis to solve the cascade equations.
The analytical expressions under Approximation A obtained by Rossi
and Greisen (1941) for the differential spectra are implemented
here. At energies $E\gg m_e c^2 Z^{-1/3}/\alpha$ collision processes and Compton effect can be neglected and the asymptotic formulae used to describe radiation processes and pair production. The average behavior of the cascade under these
assumptions can be completely solved, and the results obtained are
in good agreement with more recent numerical solutions obtained by Aharonian \& Plyasheshnikov (2003).

The natural unit of thickness in shower theory is the radiation
length $X_0$, given by
\begin{equation}
\frac{1}{X_0}=n\,\left(2.318\,Z\,(Z+1)\frac{\ln
(183\,Z^{-1/3})}{1+0.12(Z/82)^2}\,\,{\rm mb}\right), \end{equation}
where $Z$ is the charge of the target nucleus and $n$ is the number
density. The number of the injected photons of energy
$E_0=E_\gamma^{\rm max}$ that will develop cascades inside the
disk can be estimated by $N_{\rm inj}=q_\gamma\,n_p\pi R_0^2
\lambda_{pp} (1-e^{-2h/\lambda_{pp}})E_0$, resulting in $\simeq
9.7.\times 10^{34}$ photons of energy $E_0\sim 830$ TeV per second  at
$t=0$, and $N_{\rm inj}\simeq 3.5\times 10^{34}$ injected photons at
the beginning of the outburst ($t=-11$ days) with $E_0\sim 215$ TeV.

As it can be already noted from the optical depth $\tau_{\gamma p}$
in Figure \ref{inicio-cas}, the cascade traverses a thicker disk
at the maximum of the outburst than at its beginning. This fact is
also emphasized in Figure \ref{aprox-A} where the end of the disk
is indicated. The cascade traverses $100$ and $22.8$ radiation lengths, respectively, for $t=0$ and $t=-11$ days, where the unit of thickness is $X_0\sim3.4\times10^4$ cm and $ 17.8\times10^4$ cm, respectively. The difference in grammage leads
to the differences in the resulting spectra for the outgoing
$\gamma$-ray photons, shown in the same figure.

On the other hand, at the beginning of the outburst the emitted luminosity
after the cascades develop in the disk is $E^2
dn/dE\sim 2\times10^{28}\,(E/{\rm GeV})^{-0.6}$ erg s$^{-1}$. This
contribution is added to the unaffected outgoing luminosity $L_{\rm
esc}^0\sim 2\times 10^{23} (E/{\rm GeV})^2$ erg  s$^{-1}$, which dominates 
at energies above $\sim 40$ GeV.

\begin{figure*}
\centering \psfig{figure=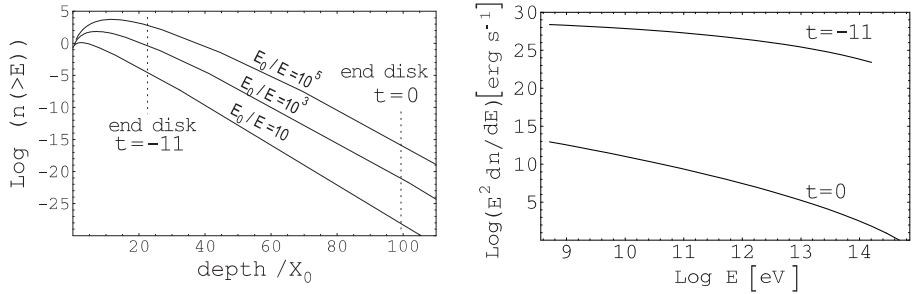, width=12.0cm}
\caption{Results for the averaged behavior of the cascades in the disk initiated by photons. Left: Number of photons of energy greater than $E$ in a shower initiated by one photon of energy $E_0$. The thickness of the accretion disk varies with time, and here this is indicated for two representative extreme times of the X-ray outburst: its beginning $(t=-11)$, and the moment of maximum intensity $(t=0)$. Right: The luminosity that emerges at the end of the disk after the development of the cascades, calculated for the most energetic photons injected by $\pi^0$ decays ($E_0=215$ TeV for $t=-11$ and $E_0=828$ TeV for $t=0$). Units of time are always in days measured from the peak of the X-ray ligth curve, luminosities are in erg s$^{-1}$}\label{aprox-A}
\end{figure*}

\section{Photospheric opacity: Inverse Compton cascades}
The opacity to propagation of $\gamma$-ray photons after their generation in the disk is determined by a rather complex combination of terms. At some GeVs the absorption will be dominated by the X-ray photospheric components. They are the emission by the polar caps of the NS (hard X-rays) and the accretion disk (soft X-rays). See Orellana \& Romero (2005) for rough estimates of the optical depth $\tau_{\gamma X}$.

Higher energy photons will interact
with the photons emitted by the massive stellar companion (Bednarek, 1997, 2000, 2006). The  stellar wind material can have a temperature $\sim 10^4$ K close to the star, adding a photon field component that effectively  absorbs $\gamma$-rays with TeV energies (Bosch-Ramon et al. 2006b, Orellana \& Romero 2006), though here we do not consider such a contribution. The pairs created by absorption can boost lower-energy photons to very high energy via Inverse Compton scattering, and the absorption of these new $\gamma$-rays leads to pair creation, so electromagnetic cascades are developed.

We have computed the high-energy $\gamma$-ray spectra formed in cascades
traversing the anisotropic stellar radiation field. Monte Carlo simulations are performed
after developing a computational code based on the scheme outlined by
Protheroe (1986) and Protheroe et al. (1992). The applied techniques enable the use of exact
cross-sections for all the relevant processes. The IC spectra are calculated as in Jones (1968), modified in similar way to Bednarek (1997).

We follow the cascade initiated by the injected flux
$\dot{n}(E)\propto E$ photons per second that escapes from the
disk at the beginning of the outburst, having  energies greater than 40 GeV (i.e. the component reprocessed by matter cascades is not considered here).
The absorber radiation field has a black-body spectrum. It comes from the B-type star located at a distance
$a\simeq 10^{13}$ cm from the neutron star, and characterized by $T_{\rm eff}\sim
30000$ K and $R_\star=10^{12}$ cm. 

We introduce the effects of the finite size of the star and the spatial variation of the field density by considering the geometric configuration as in Dubus (2006). At distances greater than $\sim 7$R$_\star$ the star is considered as a point source, whose photon density profile decays with the square distance. For simplicity, effects of the presence of a magnetic field are disregarded in these simulations,
though it may influence the pure IC cascade by changing the
lepton direction of propagation (Bednarek, 1997).

\begin{figure}
\centering \psfig{figure=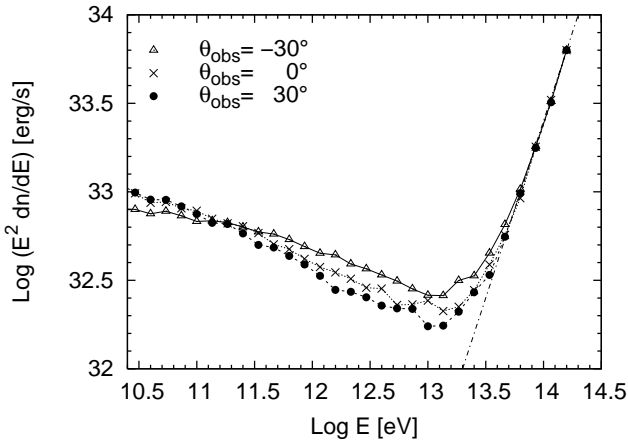, width=6.0cm,angle=-90}
\caption{Resulting luminosity from photospheric IC cascades, for different
values of the observing angle. The slope of the injected spectrum is indicated
by the line.}\label{cascadaIC}
\end{figure}

 The distance from the
source to where we follow the cascades is of $\sim 10$ times the orbital separation. Specifically, photons that go beyond $\Delta x\sim 10\,a$ are stored to form the outgoing spectra. Further away, the cascade efficiency is reduced by the dilution of the background radiation field. The emerging
secondary $\gamma$-rays depend on the angle of their escape
from the accretion disk, $\theta_{\rm obs}$ (see the sketch in
Fig. \ref{cheng}). This angle is measured from the $z$ direction toward
the B-type star in the plane that contains both the primary and the NS.
Some resulting spectra are shown in Figure
\ref{cascadaIC}, for different values of $\theta_{\rm
obs}$. These spectra are obtained by sorting
photons into bins of width $\Delta (\log E)=0.15$, and imposing energy conservation to obtain the position of the curve in the energy-luminosity plane. Note that the more energetic part of the spectra is not affected by the cascades, so the photons of energy $E_\gamma^{\rm max}$ still dominate the total luminosity.

According to our results, the spectra of escaping cascade $\gamma$-rays present only a slight dependence on the angle of propagation, at energies above $\sim 10$ GeV. The general tendency of the results is in accordance with the expectation of a deeper absorption closer to the star. It is clear that the cascade effect have led to results very different from those expected by just the incorporation of a factor $\exp({-\tau})$. 
The effect of the injection of a monoenergetic beam of protons in the disk is the final production of a power-law gamma-ray emission of photon index $\sim -2.2$, between $\sim 10$ GeV and $\sim 10$ TeV. The luminosities obtained are about $6\times 10^{32}$ erg s$^{-1}$ at $100$ GeV. The spectral features shown in the Figure \ref{cascadaIC} should be detectable by a Cherenkov telescope like MAGIC (A0535+26 is not in the field of view of HESS) if combined efforts with the X-ray observers are successful, in the sense that a the high-energy observations should be made at the beginning or ending of a major X-ray outburst. 
 

\section{Discussion and summary}

We revisited the scenario of the high energy
gamma-ray emission in an accreting magnetized neutron star that forms
part of a HMXB. The evolution of the physical parameters during a
major X-ray outburst was calculated for A0535+26. The results obtained point to an anti-correlation between the X-ray and the TeV fluxes. 

The energy power injected in the disk by the proton bombardment
($\dot{E}_{\rm inj}$) should perturb its structure. 
Under perturbations, the standard disk model can develop thermal instabilities (Shakura \& Sunyaev, 1976). If the perturbation affects the radiation pressure dominated zone, it will swell up exponentially, with a time scale in the range of seconds to minutes for an accreting NS. The problem
of the stability of a standard disk is currently treated with numerical
simulations. Recent improvements on the topic were made by Teresi
et al. (2004), and by Merloni \& Nayakshin (2006).
According to them, taking into account the stabilizing effect of
radial advection, a limit cycle type of behavior should be
expected, probably giving rise to QPO features in the X-ray signal. 
The perturbing effect of proton bombardment
on the disk should result in an oscillation of the disk height with moderate amplitude, as those studied by Teresi et al. (2004). If this does not happen, we speculate that the alternative effect of proton bombardment could be to
 inflate the disk enough to stop the mechanism responsible of the perturbation (i.e. by filling the
accelerating gap). Without the proton injection, the disk deflates and eventually a kind of periodic behavior could also be established.

We have performed Monte Carlo simulations of the electromagnetic cascades initiated in the photosphere of the system by the high energy $\gamma$-ray photons that emerge unabsorbed from the disk. 
The outgoing spectrum is characterized by a power law extending from $\sim 10$ GeV up to $\sim 10$ TeV. At higher energies the original spectrum is not affected by cascades.
The calculations presented here can be improved through more specific geometric
considerations, and the inclusion of the effect of a random magnetic field component. 

Luminosity levels predicted at the TeV domain of $\sim 10^{33}$ erg s$^{-1}$, coming from galactic sources similar to A0535+26 at the instances of the beginning or the ending of a major X-ray outburst, should be detectable by the new generation of Cherenkov telescopes. The spectral features imprinted by the photospheric opacity could be detected as well. Future observation of high-energy $\gamma$-ray photons emitted by HMXBs with accreting magnetized pulsars should be planned in quick response to alerts of renewed X-ray activity in the sources.

\begin{acknowledgements}
We thank an anonimous referee for very constructive comments.
We benefited from valuable discussions with V. Bosch-Ramon and J. M. Paredes. 
M.O. would like to thank Paula Benaglia for kindly sharing her
computer during first steps of programing. This research has been
supported by CONICET (PIP 5375) and the Argentine agency ANPCyT through Grant
PICT 03-13291 BID 1728/OC-AR.

\end{acknowledgements}
{}

\begin{thebibliography}{}
\bibitem{} Aharonian, F. A., et al. (HESS coll.) 2005a, Science, 309, 746
\bibitem{} Aharonian, F. A., et al. (HESS coll.) 2005b, A\&A, 442, 1  
\bibitem{} Aharonian, F. A., Atoyan, A. M. 1981, Ap\&SS, 79, 321
\bibitem{} Aharonian, F. A., Plyasheshnikov, A. V. 2003,
Astrop. Phys., 191, 525
\bibitem{} Albert, J. et al. 2006, Science, 312, 1771
\bibitem{} Anchordoqui, L. A., et al. 2003, ApJ, 589, 481
\bibitem{} Bednarek, W. 1997, A\&A 322, 523
\bibitem{} Bednarek, W. 2000, A\&A 362,646
\bibitem{} Bednarek, W. 2006, MNRAS, 368, 579 
\bibitem{} Begelman, M.C., Rudak, B., Sikora, M. 1990, ApJ, 362, 38
\bibitem{} Bj\"ornsson, G., et al. 1996, ApJ, 467, 99
\bibitem{}Bosch-Ramon, V., Romero, G.E., Paredes, J.M. 2006a, A\&A, 447, 263 
\bibitem{}Bosch-Ramon, V. et al. 2006b, submitted to A\&A,   
\bibitem{} Coe, M.J., et al. 2006 MNRAS, 368, 447
\bibitem{} Cheng, K. S., et al. 1992, J. Phys. G.: Nuc. Part. Phys. 18, 725
\bibitem{} Cheng, K. S., Ruderman, M. 1989, ApJ, 337, L77
\bibitem{} Cheng, K. S., Ruderman, M. 1991, ApJ, 373, 187
\bibitem{} Cheng, K. S., et al. 1991, ApJ, 379, 290 
\bibitem{}Dubus, W. 2006, A\&A,  451, 9 
\bibitem{} Finger, M.H., Wilson, R.B., Harmon, A.B. 1996, ApJ 459, 288
\bibitem{} Gosh, P., Lamb, F.K. 1979, ApJ, 234, 296 
\bibitem{} Giovannelli, F., Graziati, L. S. 1992, Sp. Sci. Rev., 59, 1
\bibitem{} Haigh, N. J., Coe, M. J., Fabregat, J. 2004 MNRAS, 350, 1457
\bibitem{}Jones, F., Physical Review 1968, 167, 1159 
\bibitem[1997]{Kni97}
Kniffen, D.~A., et al. 1997, ApJ, 486, 126.
Liu, Q.~Z., van Paradijs, J., van den Heuvel, E.~P.~J. 2000,
A\&AS, 147, 25.
\bibitem{} Merloni, A., Nayakshin, S. 2006, MNRAS, 372, 728
\bibitem{}Mori, M., et al. 1997, ApJ 476, 842
\bibitem{} Orellana, M., Romero, G. E. 2005, Ap\&SS, 297, 167
\bibitem{} Orellana, M., Romero, G. E. 2006, Ap\&SS, in press [astro-ph/0608707]
\bibitem[2000]{Par00}
Paredes, J.~M., Mart\'{\i}, J., Rib\'o, M., Massi, M. 2000,
Science, 288, 2340.
\bibitem{} Pfrommer, C., En$\beta$lin, T.A. 2004 A\&A, 413,17
\bibitem{} Protheroe, R. J. 1986, MNRAS, 221, 769
\bibitem{} Protheroe, R. J.,Mastichiadis, A., Dermer, C.D. 1992 Astrop. Phys., 1, 113
\bibitem{} Romero, G.E., et al. 1999, A\&A, 348, 868
\bibitem{} Romero, G.E., et al. 2001, A\&A , 376, 599
\bibitem{} Romero, G.E., et al. 2003, A\&A, 410, L1
\bibitem{} Rossi B., Greisen, K. 1941 Reviews on Modern Physics 13, 240
\bibitem{} Rossi B. 1952, {\em High Energy Particles}, Prentice-Hall, Englewood Cliffs
\bibitem{} Rudak, B.  M\'esz\'aros, P. 1991, ApJ 383, 269
\bibitem{} Shakura, N. I., Sunyaev, R. A. 1976, MNRAS, 175, 613
\bibitem{} Sierpowska, A., Bednarek, W. 2005, MNRAS, 356, 711
\bibitem{} Stella, L., White, N.E., Rosner, R. 1986, ApJ, 308, 669
\bibitem{} Teresi, V., Molteni, D., Toscano, E. 2004, MNRAS, 351, 297
\bibitem{wang} Wang, Y.M. 1996, ApJ, 465, L111

\end{thebibliography}
\end{document}